\documentclass[a4paper]{jpconf}

\usepackage{graphicx}

\begin{document}
\title{Gravothermal instability with a cosmological constant in the canonical ensemble\footnote{Text of a talk given at the workshop \textit{Gravity, Quantum and Black Holes} in Budapest, in September 2012. Details and extensions of the results, reported briefly here, will appear elsewhere.}}

\author{Minos Axenides$^1$, George Georgiou$^1$ and Zacharias Roupas$^{1,2}$}
\address{$^1$Institute of Nuclear and Particle Physics, N.C.S.R. Demokritos, GR-15310 Athens, Greece} 
\address{$^2$Physics Department, National Technical University of Athens, GR-15780, Athens, Greece}
\ead{roupas@inp.demokritos.gr}

\begin{abstract}
We present here how the gravothermal or Antonov's instability, which was originally formulated in the microcanonical ensemble, is modified in the presence of a cosmological constant and in the canonical ensemble. In contrast to the microcanonical ensemble, there is a minimum, and not maximum, radius for which metastable states exist. In addition this critical radius is decreasing, and not increasing, with increasing cosmological constant. The minimum temperature for which metastable states exist is decreasing with increasing cosmological constant, while above some positive value of the cosmological constant, there appears a second critical temperature. For lower temperatures than the second critical temperature value, metastable states reappear, indicating a typical reentrant phase transition. The two critical temperatures merge when the cosmological density equals one half the mean density of the system.
\end{abstract}

\section{Introduction}
\indent Antonov in 1962 \cite{Antonov} discovered an instability that later became known as Antonov's instability or gravothermal catastrophe \cite{Bell-Wood}. It signified the beginning of the study of thermodynamics of self-gravitating systems \cite{Padman,Chavanis,Katz,Chavanis2} and the nowadays rapidly developing field of statistical mechanics with long-range interactions \cite{Bell, Dauxois}. \\
\indent In the original formulation of Antonov's instability \cite{Antonov,Bell-Wood}, the system consists of particles (stars) bound by a spherical shell with insulating and perfectly reflecting walls. The particles interact only with Newtonian gravity (no cosmological constant or any relativistic effects are present) and the number of particles and the energy of the system are constant, i.e. the system is studied in the microcanonical ensemble. It was found that the entropy of the system has no global maximum and that for $ER/GM^2 < -0.335$ no equilibrium states exist. For $ER/GM^2 > -0.335$ there exist local entropy maxima, i.e. metastable states, only for density contrast below the critical value $(\rho_0/\rho_R)_{cr} = 709$, where $\rho_0$, $\rho_R$ are the central and edge densities, respectively. \\
\indent In gravity the ensembles are not equivalent. In the microcanonical ensemble, stable equilibria with negative specific heat do exist. This negative specific heat region is replaced by a phase transition in the canonical ensemble \cite{Padman}. The Antonov system in the canonical ensemble (with no cosmological constant),  is studied in detail by Chavanis \cite{Chavanis}. He found that the free energy does not
have a global minimum and equilibria exist only for temperatures greater than the critical dimensionless temperature $(RT/GM)_{cr} = 0.40$ which corresponds to dimensionless inverse temperature $(GM\beta/R)_{cr} = 2.52$. These equilibria are stable only for density contrast less than the critical value $(\rho_0/\rho_R)_{cr} = 32.1$ (see Ref. \cite{Katz2}, as well). \\
\indent We studied the gravothermal instability with a cosmological constant $\Lambda$ in the microcanonical ensemble in Ref. \cite{we}. It was found that for a fixed energy, the maximum radius (which equals $R_{cr} = (-0.335/E)GM^2$ with no $\Lambda$) is increasing with increasing $\Lambda$, while for $\Lambda > 0$ there appears a second critical radius which is decreasing and the two radii merge at some marginal $\Lambda$. This was identified as a reentrant behaviour (see Ref. 
\cite{reentrant1,reentrant2,reentrant3,reentrant4} for details on reentrant phase transitions in statistical systems with long-range interactions). In the present work, the system is placed in a heat bath and the walls are non-insulating, i.e. we study the canonical ensemble. In contrast to the microcanonical one, we find that the critical radius $R_{cr}$, for a fixed temperature and mass, is decreasing with increasing cosmological constant. In addition the instability occurs for $R<R_{cr}$ in the canonical ensemble, in contrast to the microcanonical one. This is evident in case $\Lambda = 0$, since then there exist no equilibria if $GM\beta/R > 2.52 \Rightarrow R < GM\beta/2.52$ for a fixed temperature and mass. A reentrant phase transition appears in the canonical ensemble, when one examines the system with respect to the temperature for a fixed radius. There appear two critical temperatures for $\Lambda > 0$. For the intermediate temperature values, no thermodynamic equilibria are possible.

\section{Analysis and results}
\begin{figure}[t]
\begin{minipage}{18pc}
\includegraphics[width=18pc]{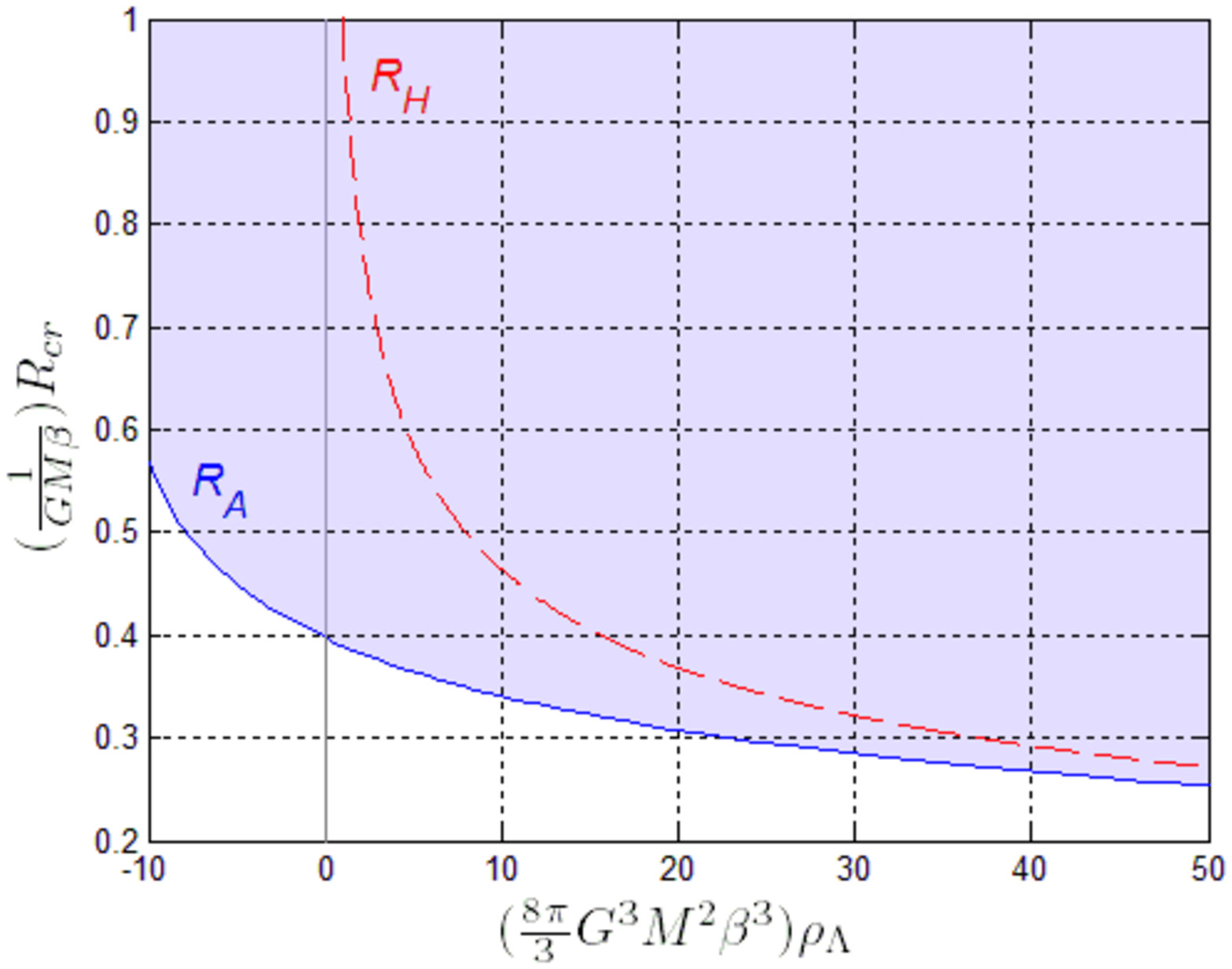} 
\caption{\label{fig:Rcr}The critical radius w.r.t. $\rho_\Lambda$ for fixed $M$, $\beta$.}
\end{minipage}\hspace{2pc}%
\begin{minipage}{18pc}
\includegraphics[width=18pc]{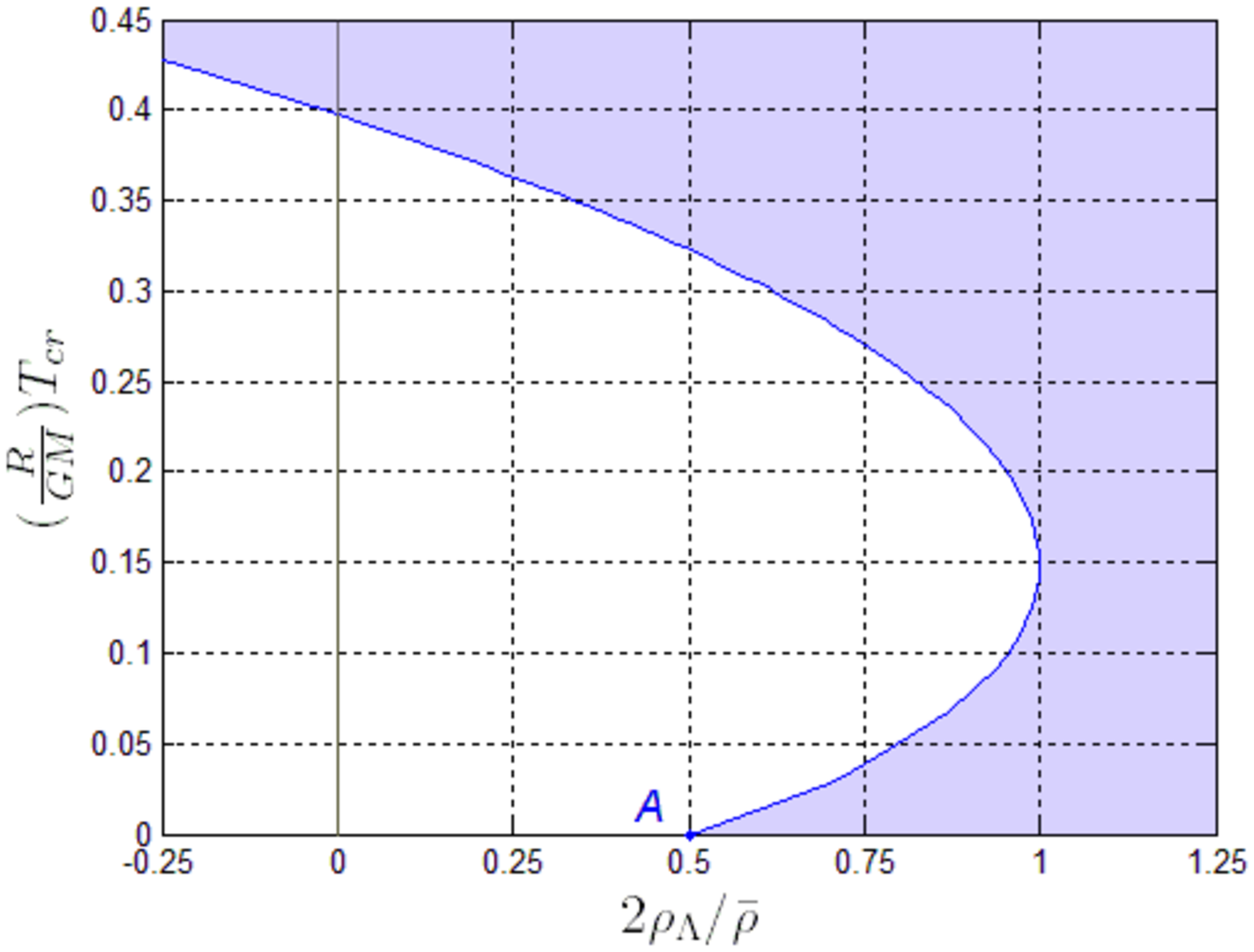}
\caption{\label{fig:Tcr}The critical temperature w.r.t. $\rho_\Lambda$ for fixed $M$, $R$.}
\end{minipage} 
\end{figure}

\indent In the Newtonian limit, the Poisson equation is modified \cite{Wald,Axenides} in the presence of a cosmological constant $\Lambda$ as:
\begin{equation}\label{eq:PoisL}
	\nabla^2 \phi = 4\pi G \rho - 8\pi G \rho_\Lambda
\end{equation}
with $\rho_\Lambda = \frac{\Lambda c^2}{8\pi G}$. Thus, the gravitational potential can be decomposed to two parts
\begin{equation}
	\phi = \phi_N + \phi_\Lambda
\end{equation} 
where $\phi_N$ is the Newtonian and $\phi_\Lambda$ the cosmological potential, that are given by
\begin{equation}\label{eq:potentials}
		\phi_N = -G\int{\frac{\rho(r')}{|\vec{r}-\vec{r}\, '|}d^3\vec{r}\, '} \; , \;
	\phi_\Lambda = -\frac{4\pi G}{3}\rho_\Lambda r^2	
\end{equation}
For convenience we call the Newtonian limit of de Sitter ($\rho_\Lambda>0$) and Anti-de Sitter ($\rho_\Lambda < 0$) spaces \cite{NewtonHook} just dS and AdS, respectively. \\
\indent We consider a self-gravitating gas of $N$ particles with unity mass inside a spherical non-insulating shell inside a heat bath and restrict only to spherical symmetric configurations. We work in the mean field approximation, for which the $N$-body distribution function is replaced by the one body distribution function $f(\vec{r},\vec{\upsilon})$. The Boltzmann entropy is defined as $S = -k\int f\ln f d^3\vec{r} d^3\vec{p}$. The Helmholtz free energy is equal to $F = E - TS$. Equivalently one can work with the Massieu function \cite{Chavanis,Katz2} $J = -F/T$ that gives
\begin{equation}
	J = S-\frac{1}{T}E
\end{equation} 
The maximization of $J$ with constant $T$ and the maximization of $S$ with constant $E$ (and constant $M$ in both cases), with respect to perturbations $\delta\rho$, are the same \cite{Chavanis} to first order in $\delta\rho$ , i.e. give the same equilibria, described by the Maxwell-Boltzmann distribution function
\begin{equation}
	f = \left( \frac{\beta}{2\pi}\right)^{\frac{3}{2}}\rho(r)e^{-\frac{1}{2}\beta\upsilon^2}
\end{equation}
where
\begin{equation}\label{eq:rho}
	\rho(r) = \rho_0 e^{-\beta(\phi - \phi(0))}
\end{equation}
This is proved in \cite{Chavanis} without a cosmological constant, and holds in the presence of $\Lambda$ as well, with the difference being that $\phi$ does not satisfy Poisson equation now, but equation (\ref{eq:PoisL}). The two ensembles have the same equilibria. However, the second variations of $J$ and $S$ are different. Therefore, what is different in the two ensembles is the turning point at which an instability sets in. This causes a great qualitative difference for the two ensembles. \\
\indent Introducing the dimensionless variables $y = \beta(\phi - \phi (0))$, $x = r\sqrt{4\pi G \rho_0\beta}$ and $\lambda = 2\rho_\Lambda/\rho_0$, and using equation (\ref{eq:rho}), equation (\ref{eq:PoisL}) becomes
\begin{equation}\label{eq:emdenL}
	\frac{1}{x^2}\frac{d}{dx}\left( x^2\frac{d}{dx}y\right) = e^{-y} - \lambda
\end{equation}
called the Emden-$\Lambda$ equation. Let us call $z = R\sqrt{4\pi G \rho_0\beta}$ the value of $x$ at $R$. In order to generate the series of equilibria needed to study the stability of the system, the Emden-$\Lambda$ equation has to be solved with initial conditions $y(0)=y'(0)=0$, keeping $M$ constant and for various values of the parameters $\rho_\Lambda$, $\beta$, $\rho_0$. This is a rather complicated problem, since, unlike $\Lambda = 0$ case, while solving for various $z$, mass is not automatically preserved, because of the mass scale $M_\Lambda = \rho_\Lambda\frac{4}{3}\pi R^3$ that $\Lambda$ introduces. A suitable $\lambda$ value has to be chosen at each $z$.
We define the dimensionless mass 
\begin{equation}\label{eq:mDEF}
	m \equiv \frac{M}{2M_\Lambda} = \frac{3}{8\pi}\frac{M}{\rho_\Lambda R^3} = \frac{\bar{\rho}}{2\rho_\Lambda}
\end{equation}
where $\bar{\rho}$ is the mean density of matter. Calling $z = R\sqrt{4\pi G \rho_0\beta}$ the value of $x$ at $R$, equation (\ref{eq:mDEF}) can also be written as $m = 3 B/\lambda z^2$
where $B = GM\beta/R$ is the dimensionless inverse temperature. It can be calculated by integrating the Emden-$\Lambda$ equation, to get:
\begin{equation}\label{eq:beta}
	B(z) = z y'(z) + \frac{1}{3}\lambda z^2
\end{equation}
We developed an algorithm to solve equation (\ref{eq:emdenL}) for various values of $\lambda$, $z$ keeping $m$ fixed.
From equation (\ref{eq:mDEF}) it is clear that solving for various fixed $m$ can be interpreted as solving for various $\rho_\Lambda$ and/or $R$ for a fixed $M$. \\
\indent We find that in dS case there exist multiple series of equilibria for a given cosmological constant, while in AdS case there is just one series likewise $\Lambda = 0$ case. Studying the series $\beta(\frac{\rho_0}{\rho_R})$ for both AdS and dS, we determined the critical values of radii and temperature beyond which there are no equilibria. These critical values are plotted in Figures \ref{fig:Rcr} and \ref{fig:Tcr}, with respect to $\rho_\Lambda$, where the unshaded region is the region of instability in both figures. \\
\indent In Figure \ref{fig:Rcr} the radius denoted $R_A$ is the minimum radius for which thermodynamic equilibria exist in the canonical ensemble. $R_H$ is the radius of the homogeneous solution which exists only in dS and is defined by equation $\rho = 2\rho_\Lambda = const$, i.e.
$
	R_H = (3M/(8\pi\rho_\Lambda))^\frac{1}{3}
$.
In the shaded region of Figure \ref{fig:Rcr} there can always be found metastable states. For AdS case ($\rho_\Lambda < 0$), these have always monotonically decreasing density ($\rho_0 > \rho_R$) and suffer a transition to unstable equilibria at a critical density contrast $(\rho_0/\rho_R)_{cr}$ whose value depends on $\rho_\Lambda$. For dS case and for $R < R_H$, the metastable states suffer a transition to unstable equilibria likewise AdS, but the density is not necessarily monotonic, neither does $\rho_0 > \rho_R$ always hold. For each $\rho_\Lambda$ there exists a tower of solutions with qualitative (and quantitative of course) different density functions. In the region $R>R_H$, all metastable states have $\rho_0 < \rho_R$. The ones, whose density is a monotonic function of $R$ do not suffer any transition to unstable equilibria, while all others series have a turning point. \\
\indent In Figure \ref{fig:Tcr} we see that in AdS the critical temperature is decreasing for increasing cosmological constant, while in dS there appears a reentrant phase transition. For small positive values of $\rho_\Lambda$, compared to the mean density $\bar{\rho}$, the system cannot stabilize for low temperatures. It needs higher temperature values to obtain a significant pressure gradient to balance gravity. As $\rho_\Lambda$ is increasing, the cosmological repelling force increases, enforcing the total outward pushing force and hence, enabling the system to stabilize at lower temperatures. Point $A$, denotes the marginal value $\rho_\Lambda^{min} = \bar{\rho}/4$ for which a dynamically static equilibrium is allowed. This limiting configuration corresponds to a singular solution at which all mass is concentrated at the edge. It can easily be calculated: $GM/2R^2  = \frac{8\pi G}{3}\rho_\Lambda^{\min} R$ $\Rightarrow$ $\rho_\Lambda^{min} = 3M/16\pi R^3$ $=$ $\bar{\rho}/4$. For $\rho_\Lambda > \rho_\Lambda^{min}$, solutions with $T=0$ (static dynamical equilibria) do exist and as $\rho_\Lambda$ increases the mass is allowed to be closer to the center. All thermodynamic equilibria below the lower critical branch (low temperatures) have increasing density ($\rho_0 < \rho_R$). For fixed $R$, $M$ and a fixed $\rho_\Lambda$ with $\bar{\rho}/4 < \rho_\Lambda < \bar{\rho}/2$, as we increase the temperature beginning from zero, we pass from a region of metastable states with $\rho_0 < \rho_R$ to a region of no equilibria, at a first critical temperature $T_1$. We have to increase the temperature even more to reach a new region of metastable states at a second critical temperature $T_2$. For $T > T_2$ the metastable states have decreasing density $\rho_0 > \rho_R$ if $R<R_H$. This is a reentrant phase transition similar to the ones found in other statistical systems with long-range interactions \cite{reentrant1}.

\end{document}